\documentstyle[12pt]{article}

\begin{document}

\centerline {\Large {Photon redshift and the appearance of a naked singularity}}

\hfill\break
\hfill\break

\centerline{{\bf I. H. Dwivedi}}\hfill\break
\centerline{{\bf Tata Institute of Fundamental Research}}\hfill\break
\centerline{{\bf Homi Bhabha Road, Bombay 400 005}}\hfill\break
\centerline{{\bf India}}\hfill\break

\vspace{1in}

\centerline {\bf {ABSTRACT}}

In this paper we analyze the redshift as observed by an external
observer receiving photons which terminate in the past at the naked 
singularity formed in a Tolman-Bondi dust collapse. Within the context
of models considered here it is shown that photons emitted from a 
weak curvature naked singularity are always finitely redshifted to an 
external observer. Certain cases of strong curvature naked singularities, 
including the self-similar one, where the photons are infinitely 
redshifted are also pointed out.

\vfil\eject

\noindent
{\bf I. INTRODUCTION}

It is generally believed that star with sufficient
mass undergoing a gravitational collapse would release very high energy 
during the collapse process. For an external observer, gravitational
collapse of a spherically symmetric star, appears to take an infinite time 
(observer time $T_o$) to reach its Schwarzschild radius $2M$.
Theoretically therefore a collapsing star should be visible for ever. 
The redshift$^1$ $z$ (frequency shift)  in late stages of collapse, however, 
\begin{equation} 1+z\propto\exp{({t_o\over 4M})}\end{equation}
increases exponentially with a very short characteristic
time( and subsequently the luminosity decays exponentially with a
short characteristic time) and hence a collapsing star would rather 
disappear in a short period of time. The end state of such a gravitational 
collapse could either be a naked singularity or a blackhole. The end 
result of gravitational collapse is rather crucial to cosmic censorship 
according to which a gravitational collapse must necessarily end in a 
singularity that is covered i.e. no light rays can escape to an external 
observer from the singularity thus formed. Recently many examples
of naked singularities have been given$^{2-6}$. We can categorize these naked 
singularity examples into basically two categories. In the first kind 
the naked singularity has only a single future directed radial null geodesic 
terminating at the singularity in the past while in second there is a 
family of radial null geodesics terminating at the naked singularity 
in the past. Therefore while the former kind of the naked singularity would 
only be visible for a instant  the later one could be visible to an external 
observer for ever. The redshift factor therefore becomes very important 
because if the photons escaping from the naked singularity
are infinitely redshifted as calculations for certain very few models so far 
suggest$^{3,4}$, it could be argued that these
examples of naked singularities though theoretically admissible
in general relativity would, however, not be visible to external observer. 

The aim in this paper is therefore to analyze the redshift
from this perspective. Tolman-Bondi dust
collapse models have been investigated quite extensively in this
context. It has been shown$^{3,6}$ that wide classes of solutions within the
Tolman-Bondi spacetime do allow the formation of both strong curvature
and weak curvature naked singularities of
either categories as mentioned above. We investigate these dust collapse
models for redshift analysis. We do find that for self-similar
dust models, photons are infinitely redshifted as shown in earlier work$^4$.
For the dust models considered here and where the naked singularity is
a strong curvature singularity, the redshift is also infinite.
However, for dust models where the singularity is a weak curvature singularity
$^9$
suitably characterized by the mass function $F(r)$
and the energy function $f(r)$,  the redshift  is finite. 
Strong curvature naked singularities$9$ are thus physically 
censored in
the sense that no amount of energy would escape from the singularity
on account of infinite redshift, the weak curvature naked singularity
on the other hand have a finite redshift and may be instantaneously 
visible. For globally naked weak curvature singularities
the redshift, for photons terminating in the past
at the naked singularity and crossing the boundary of the star at late
stages near Schwarzschild radius to reach the distant observer, 
increases exponentially but with a different time constant compared to
the photons emitted from the surface of the star. Thus it is likely that
luminescence of the naked singularity and that of the surface of the star
would not disappear simultaneously in such cases.

The redshift factor is defined as follows.
Consider a source with 4-velocity $u^a_{(s)}$ and an observer with 4-velocity
$u^a_{(o)}$  located at events $P_1$ and $P_2$ in an arbitrary spacetime. Let
$k^a\equiv dx^a/d\lambda$ be the tangent vector to the null geodesic 
connecting the two events $P_1$ and $P_2$ with $\lambda$ being the affine 
parameter. The redshift $z$ (i.e. frequency shift) is defined as
\begin{equation} 1+z={[k_au^a_{(s)}]_{P_1}\over [k_au^a_{(o)}]_{P_2}}
\end{equation}
where the numerator and the denominator are evaluated at the events
$P_1$ and $P_2$ respectively which denote emission and reception of light.
In the above relation  the source or the observer need not be in the
same coordinate patch. Thus the source for example could be in the interior 
dust cloud described by the  Tolman-Bondi solutions, while the observer could 
be a Schwarzschild observer in the external Schwarzschild spacetime. However, 
at the boundary the geodesics connecting the two events must be continuous.

In order to calculate the redshift, besides having the expressions
for $u^a_{(s)}, u^a_{(o)}$ one must have the trajectory of photons 
as well as the tangent $k^a$ to the null geodesic connecting the 
event $P_1$ at the source and $P_2$ at the observer
and one must evaluate them as per 
equation (2) at the event $P_1$ when the light emitted and at $P_2$ when it
is received. To do that we therefore discuss below Tolman-Bondi
dust models and escaping photons from singularities in such a collapse.

\noindent
{\bf II. Tolman-Bondi Dust collapse}

Spherically symmetric dust collapse in the comoving coordinates 
is given by the Tolman-Bondi metric$^7$
\begin{equation}ds^2= -dt^2+{R'^2\over1+f}dr^2+R^2(d\theta^2+sin^2
\theta d\phi^2)\end{equation}
where $R=R(t,r)$ is interpreted as radius of the spherical shells in the
sense that $4\pi R^2(t,r)$ gives the proper area of shells and is given by
\begin{equation}\dot R=-\sqrt{{F\over R}+f}\end{equation}
The dot and prime denote partial derivatives with respect
to the coordinates $t$ and $r$ respectively. Arbitrary functions $F(r)\ge 0$ 
and $f(r)$ are atleast $C^2$ functions of coordinate $r$ throughout the
dust cloud and are interpreted as mass and energy
functions respectively. The collapse leads to the formation of a shell
focusing singularity at points described by the curve $t=t_o(r)$ 
such that $R(t_o(r),r)=0$. The singularity at $r=0$ is the central 
singularity.
Only central singularity could be naked while singularities at $r>0$ are
necessarily covered. The singularity at $r=0$ occurs at time $t=t_o(0)=t_s$
such that $R(t_s,0)=0$. Energy conditions are required to be satisfied
throughout the spacetime and we would consider only the cases where 
no shell crossing singularities occur i.e. $R'>0$ throughout the spacetime
except perhaps at the central shell focusing singularity. 
Equation (4) can be integrated with the condition
that at some initial $t=0$, $R(0,r)=r$ to yield the expression for $R(t,r)$. 
Since in this paper mostly we would be interested in only marginally 
bound cases characterized by $f(r)=0$ we would give the $R=R(t,r)$ for
marginally bound case below and
refer the reader for general expression of $R(t,r)$ to reference [6]
\begin{equation}
{2\over 3}\sqrt{{R^3\over F}}=t_o(r)-t,\quad t_o(r)={2r^{{3\over 2}}\over 3
\sqrt{F}}\end{equation}
In the context of cosmic censorship the spacetime at some initial should be 
singularity free and hence we can take the following form
of the mass function as a representative of a more general form (except
in self-similar case) as
\begin{equation}F(r)=F_or^3(1-F_1r^{\beta})\equiv F_or^3h(r)\end{equation}
where $F_0,F_1$ and $\beta \ge 1$ are positive constants. 
Note that the form $F(r)=F_o r^3 h(r)$ where $h(r)$ is at least
a $C^1$ function of $r$ is due to the requirement that at the initial
$t=0$ spacetime be nonsingular. 
First spacetime singularity for mass function given as above occurs at the 
center $r=0$ at time $t=t_s={2\over 3\sqrt{F_o}}$.
The form of $F(r)$ in (6) is quite general except as mentioned
in self-similar case where necessarily 
\begin{equation}F={\it F}_or\end{equation}
and in such a case the singularity occurs at $t=0,r=0$ while the spacetime 
is nonsingular for $t<0$. 

\noindent
{\bf III. Photons from naked singularity}

As pointed out in equation (2) the redshift factor is calculated from
source to observer along the null geodesic(photon) connecting the
source and the observer, hence the trajectory of light connecting the
source and the observer has to be known.  In present context it amounts to
null geodesics escaping from the central naked singularity and reaching the
external observer which could be either a local observer or a distant one.
Tolman-Bondi dust collapse has been extensively studied in this
context of formation of naked singularities and the escaping null geodesics
from it$^{3,6}$. It has been
shown that for wide classes of dust models characterized by
the arbitrary functions $f$ and $F$ the central singularity would be
naked. Therefore avoiding the repetition of earlier work we would
in this section only give essential expressions and results in this context
and refer the reader to reference [6] for details.

Let $k^a$ be tangent to radial null geodesics (i.e. $k^ak_a=0=k^a_{;b}k^b$)
For the metric in equation (3) we have for radial null geodesics
\begin{equation}k^r={\sqrt{1+f}\over R'}k^t\Rightarrow {dt\over dr}=
{R'\over\sqrt{1+f}}\end{equation}
\begin{equation}{dk^t\over d\lambda}=-{\dot R'\over \sqrt{1+f}}k^tk^r
\end{equation}
where $\lambda$ is an affine parameter.
First of the above equation on integration gives the trajectories
of photons as $t=t(r)$. While the later one can be integrated to
obtain the expression for $k^t$. 
Restricting ourself from now on to marginally bound case $f=0$ given by
equation (5) and following reference [6] we use $R$ and $u=r^{\alpha}$ as 
pair of variables and use (5) to write the radial null geodesic equation 
in (8) as
\begin{equation} {dR\over du}={1\over \alpha}
(1-\sqrt{{\Lambda \over X}}){H}=U(X,u)\end{equation}
\begin{equation} \Lambda={F(r)\over r^{\alpha}},\quad X={R\over r^{\alpha}},
H=H(X,r)={\eta\over 3}X+{\Theta\over \sqrt{X}}
\end{equation}
where 
\begin{equation}\alpha = 1+{2\beta\over 3},\quad \Theta= 
{\beta F_1\over 3(1-F_1r^{\beta})},\quad
\eta =3(1-\Theta r^{\beta})\end{equation}
for the mass function given in equation (6) and  
\begin{equation}\alpha = \eta =1,\quad \Theta= {2\over 3}\end{equation}
for self-similar case given by mass function in equation (7).
The integration of equation (10) gives the equation of radial null
geodesics in the form $R(t,r)=y(r)$ where $y(r)$ is some function of
$r$. The equation (10) is highly nonlinear and an exact solution
even in very simple cases is not available except for self-similar
case$^{4,6}$. We would therefore analyze the behavior of radial
null geodesics which is crucial to any calculation of redshift factor.
To find this behavior we follow the earlier work and
give the necessary discussion of outgoing radial null geodesics
from the naked singularity below and would refer the reader to reference
[6] for details.

The central singularity is at least locally naked if equation 
$V(X)\equiv U(X,0)-X=0$ has a real positive root $X=X_0$
\begin{equation}V(X)= U(X,0)-X=
\left(1-\sqrt {{\Lambda_0
\over X}}\right)
({\eta_0 \over 3}X+{\Theta_0\over \sqrt{X}})-\alpha X=0\end{equation}
where $\eta_0=\eta(0), \Theta_0=\Theta(0)$. 
In case a real positive root $X=X_o$ of the root equation (14) 
exists then there are out going radial 
null geodesics from the naked singularity and the behavior of the geodesics
in the neighborhood of the singularity is described by 
\begin{equation}R=X_0u\end{equation} 
Using equations (8) (10) the behavior in (15) of singular geodesic can be
expressed in terms of $t$ and $r$ as
\begin{equation}{t\over t_s}=1+{X_or^{\alpha}\over 1-
\sqrt{{\Lambda_o\over X_o}}}\end{equation}
Note that for the general mass function given by equation (6) 
with $\beta <3$, $V(X)=0$ has a real positive root and is given by
\begin{equation}V(X)= 
(1-\alpha)X+{\Theta_0\over\sqrt{X}}=0\Rightarrow X_o^{{3\over 2}}=
{\Theta_0\over \alpha -1}\end{equation}
For $\beta=3$ case and self similar case given by equation (7)
the root equation (14) turns out to be a biquadratic equation 
\begin{equation}V(X)=\Rightarrow 2(\sqrt{X})^4+(\sqrt{X})^3\sqrt{\Lambda_0}
-\Theta_0\sqrt{X}+\Theta_0\sqrt{\Lambda_0}=0\end{equation}
and have real positive roots if$^6$
\begin{equation}{F_1\over  F_o^{{3\over 2}}}>13 +{15\over 2}\sqrt{3}
\quad for\quad \beta =3\end{equation}
\begin{equation}{2\over  {\it F}_o^{{3\over 2}}}>13 +{15\over 2}\sqrt{3}
\quad for\quad selfsimilar\end{equation}  
respectively. Thus for the marginally bound cases considered here the 
singularity is naked and the behavior of outgoing radial null geodesics is
given by (15) in the neighborhood of the singularity.
For a external observer viewing the collapse, the singularity
would be visible if null geodesics with their past 
end point at the singularity, reach the external observer within the
the dust cloud or cross the boundary  $r=r_b$ of the star with $dR/du >0$. 
For this purpose we consider the paths of these null geodesics, 
which escape from the naked singularity with the tangent
$X=X_o$(in $(R,u)$ plane) which is the real positive root of Equation (14). 
By putting $x=X^{3/2}$, $x_0=X_0^{3/2}$, $v(x)={3\over 2}\sqrt{X}V(X)$ 
and ${\cal U}(x,u)={3\over 2}\sqrt{X} U(X,u)$ 
we have after using equations (10), (11) and (14)
\begin{equation}{dx\over du}={{\cal U}(x,u)-{\cal U}(x,0)+v(x)\over u}
={(x-x_0)(h_0-1)+S\over u}\end{equation}
where we have put
$v(x)=(x-x_0)(h_0-1)+h(x)$ such that 
$h_0-1=(dv(x)/dx)_{x=x_0}$, $h(x)$ contains
higher order terms in $(x-x_o)$
and $S=S(x,u)={\cal U}(x,u)-{\cal U}(x,0)+h(x)$ and $S(x_o,0)=0$. 
On integration of above , the null trajectories $x=x(u)$  are given by,
\begin{equation} x-x_0= Du^{h_0-1} +u^{h_0-1}\int Su^{-h_0}du\end{equation}
Here $D$ is a constant of integration that labels different geodesics$^6$.
Note that the last term in equation (22) always vanishes
due to the fact that as $u\to 0$, $x\to x_0$
regardless of the value of the
constant $h_0$. The first term on the right hand side of the equation (22) 
$Du^{h_0-1}$, however, vanishes only if $h_0>1$. 
Therefore, if $h_0>1$ a family of outgoing singular geodesics
(with each curve being labeled by different values of the constant $D$) escape
from the naked singularity with $X=X_0$ as the value of the tangent at 
the singularity in $(R,u)$ plane. On the other hand, 
in  case $h_0<1$, only single null geodesic with $D=0$ escapes from the
naked singularity with tangent $X=X_0$ in $(R,u)$ plane. 
For $\beta=3$  and the self-similar case$^{3,6}$
$h_0>1$ and hence a family of outgoing radial null geodesics escape 
from the singularity with each geodesics labeled by different value of $D$. 

For the marginally bound cases
considered here $h_0=1-(\beta /\alpha)<1$ for $\beta <3$ and therefore
only a single geodesic escapes from the naked singularity in this case
given by $D=0$. Furthermore $S(x,u)\to 0$ as $u\to 0$ for $\beta <3$. 
An important point to note in this case is that the geodesics
labeled by values of $D\ne 0$ represent radial null geodesics emitted 
from the center $r=0$ with emission time $t=t_{s0} <t_s$. In fact from
equations (5), (6) and (22) it follows as $r\to 0$
$D=[(R/r)^{3/2}]_{r=0}=1-(t_{s0}/t_s)$. $D$ vanishes for
the photon escaping from the naked central singularity at emission time 
$t_{s0}=t_s$ in such cases.

Thus radial null geodesics 
in the near regions of the naked central singularity can be described by.
\begin{equation}R^{{3\over 2}}-X_o^{{3\over 2}}u^{{3\over 2}}=
Du^{{1\over 2}+h_o} +u^{{3\over 2}}O(u^m) \quad m>0\end{equation}
where $O(u^m)\to 0$ as $u\to 0$. 

The naked central singularity would be globally
naked only if these out going radial null geodesics from the naked
singularity would cross the boundary of the star before it reaches its 
Schwarzschild radius. In many examples including the marginally bound cases 
the naked singularity is globally naked as well$^{2-6}$. 
Therefore in cases when there are a family of
null geodesics escaping from naked singularity which is globally naked as 
well, constant $D$ in equations (22) is determined by the condition
that at the boundary $u=u_b=r^{\alpha}_b$ and 
$x=(R_b/r_b^{\alpha})^{3/2}=x_b$ and we have
\begin{equation} x-x_0= (x_b-x_0)\left[{u\over u_b}\right]^{h_0-1} +
u^{h_0-1}\int_{u_b}^u Su^{-h_0}du\end{equation}
The event horizon is represented by the null geodesic for which
$x_b=\Lambda_b=(\Lambda(r_b))^{3/2}$. The null geodesics with  $x_b$  
close to $\Lambda_b$ cross the boundary of the star at late stages of collapse
when the star radius is close to $2M$ (Schwarzschild radius). 
In case of a globally naked singularity where only single radial null 
geodesics escapes from the singularity, the escaping geodesic can cross
the boundary at any stage of the collapse depending on the exact global
behavior of function $F(r)$ and the boundary of the star at $r=r_b$.
We should mention that our above discussion was limited only for
marginally bound case, however, all the above qualitative description
of outgoing radial null geodesics remain the same even for the
general cases for $f\ne 0$ and the difference comes only on the
value of $\alpha$ and in the form of equation $V(X)=0$ in (14).

\noindent
{\bf IV. Redshift}

The aim in this paper is to evaluate redshift for
light terminating at the central naked singularity in the past and 
received by an external observer. For the spacetime
described by the Tolman-Bondi metric given in equation (3) we treat
the source at $r=0$ and the observer at $r=r_o$ within the dust cloud,
Thus 4-velocities, $u^a_{(s)}$ of the source at the center 
$r=0$ and $u^a_{(o)}$ of the observer at $r=r_o$ within the dust cloud are
\begin{equation}u^a_{(s)}=\delta^a_t\end{equation}
\begin{equation}u^a_{(o)}=\delta^a_t\end{equation}
Naked central singularity at $r=0,t=t_s$ is the source of photons and is
the event $P_1$ at the source. The radial null geodesic along which the
redshift is to be evaluated are precisely the outgoing null
geodesics which terminate at the singularity in the past and reach the
external observer at event $P_2$.
The expression  $k^t$ is important from the point of calculation for
redshift and using equations (5) to (11) we get by integrating equation (9)
\begin{equation}k^t=c_o\exp{(-\int{{\dot R'\over \sqrt{1+f}}k^rd\lambda})}
\Rightarrow =c_o\exp{(-\int{N(R,r)dr})}\end{equation}
where $c_o$ is a constant and
\begin{equation}N\equiv N(R,r)={\sqrt{rF}\over 2R^2}\left(1-{\eta\over 3}
-{2\eta\over 3}
({R\over r})^{3/2}\right)\end{equation}

Using equations (25) to (28) and (2) we get the redshift factor for
photons 
\begin{equation}
1+z=\exp{(\int_{0}^{r_o}Ndr)}\end{equation}
For the general mass function given by equation (6) we can express
\begin{equation}N={\sqrt{F_0h(r)}\over 2r^{{\beta\over 3}}X^2}\left(
2(1-\Theta r^{\beta})
X^{3/2}-\Theta\right)\end{equation}
where $X=X(r)$ is to be treated as a function of $r$ describing the
path of null geodesics given by equation (22) or alternatively from equation
(23). 
The lower limit $r=0$ of the integral in equation (29) corresponds to the
naked central singularity and is also a singular point of the
integral. In order to determine the convergence of the integral let
us consider the limit of
\begin{equation}\lim_{r\to 0}[r^{{\beta\over 3}}N(X,r)]\end{equation}
As per the discussion in the previous section on the radial null geodesics
terminating at the naked central singularity at $t=t_s, r=0$ the behavior
of geodesics is given by equations (15) in the neighborhood of the
singularity and in the limit of approach to
singularity as $r\to 0, t\to t_s$ $X\to X_0$ , hence using equations
(10) to (12), (15) and (30) we get
\begin{equation}\Psi=\lim_{r\to 0}[r^{{\beta\over 3}}N(X,r)]=
{\sqrt{F_o}\over 2\sqrt{X_o}}(3-\alpha)\ne 0
\quad for\quad \beta <3\end{equation}
\begin{equation}\Psi=\lim_{r\to 0}[r^{{\beta\over 3}}N(X,r)]=\sqrt{\Lambda_0}
(X_o+{\Theta_0\over \sqrt{X_0}})>0
\quad for\quad \beta =3\end{equation}
Since $\Psi$ has a finite limiting value at the singularity for $\beta <3$
the integral converges absolutely for $\beta <3$ and therefore the
redshift given in equation (29) for such a case is finite. On the other hand
for $\beta =3$ the limiting value is finite positive hence the integral 
in equation (29) diverges.
A similar consideration in case of a selfsimilar spacetime given by
the mass function in equation (7) reveals
\begin{equation}\Psi=\lim_{r\to 0}[rN(X,r)]=
\sqrt{\Lambda_0}(X_o+{2\over 3\sqrt{X_0}})>0\end{equation}
and hence the integral diverges.

It would not be out of context to consider redshift from the point
of view of an 
external observer viewing the photons emitted from the source at center at
$r=0$ approaching the naked central singularity at $r=0$. 
The integral in equation (29)
is too complicated to evaluate exactly however we can estimate the
redshift factor and it's behavior and for the purpose of simplicity
we can take the case $\beta =1$ which corresponds to finite redshift as shown 
above. Equation (23) with constant
$D=1-{t_{so}\over t_s}$ in such a case describes the null geodesics from
the center emitted at time $t=t_{so}$. Considering a local external observer 
at $r_o<<1$, using equations (6), (12), (23), (29) and (30) and neglecting 
higher order terms in $r$, we get for the redshift
\begin{equation} 1+z\propto \exp(-3D^{{2\over 3}}
+{X_0^{{3\over2}}r_o+3D\over (X_0^{{3\over2}}r_o
+D)^{{1\over 3}}})\end{equation}
As $t_{s0}\to t_s$, $D=1-{t_{s0}\over t_s}\to 0$ and the redshift remains
finite. Thus to an external observer viewing the source at the center $r=0$ 
the redshift remains finite as the source at the center approaches the 
singularity at the center at time $t_{s0}=t_s$.

Thus we arrive at the conclusion that redshift is finite for collapse 
scenarios with $\beta <3$ and diverges for $\beta =3$ and selfsimilar cases. 
The interesting point to note that while $\beta <3$ cases correspond to a weak
curvature naked singularity the $\beta=3$ and selfsimilar cases represent
strong curvature singularities$^6$. The conclusions would be the same in
non marginally bound cases of dust collapse. Therefore while the strong 
curvature naked singularities would be physically censored owing to the 
infinite redshift
the weaker curvature singularities can have finite redshift factor. This
perhaps is the result of growth of density from outer layers of star
to the center of the star as characterized by the value of $\beta$. Thus
if the density decreases away from the center faster as characterized by
$\beta =3$ the singularity is not only a strong curvature singularity
but it also does not allow any energy to escape on the otherhand
if the decrease in the density from the center is slow i.e. $\beta <3$
the singularity is gravitationally weak and it allows the energy to escape.

\noindent
{\bf V. Schwarzschild Observer}

For a distant observer in external Schwarzschild spacetime viewing 
the late stages of a collapsing star, the redshift 
for the photons emitted
from star,s surface appears to be time dependent and increasing exponentially. Since
In the case of a globally naked singularity photons escaping from
the central naked singularity can cross the boundary of the star closer
and closer to the Schwarzschild radius. As a matter of interest we point
out very briefly as a remark on
the photon redshift for such a scenario. In the context of Tolman-Bondi
dust collapse such a scenario is possible only for examples of
globally naked central singularity where a family of null geodesics terminate
at the singularity in the past and the singularity is gravitationally weak 
as well in view of the preceding section. 
it is quite likely that there are dust models in general where the singularity
is not only weak but a family of null geodesics also terminates at the
naked singularity$^{6,8}$. 
In such cases
after first ray crosses the boundary of the star in to vacuum there
would always be geodesics escaping from the singularity to the
Schwarzschild exterior closer and closer to the apparent horizon
$R=F$ in the interior and would cross the boundary of
the star closer and closer to the Schwarzschild radius $2M$.
Let us consider a collapsing star with internal metric described by the
Tolman-Bondi dust models and the exterior of
the star is described by the vacuum Schwarzschild metric.
\begin{equation}ds^2= -(1-{2M\over r_s})dT^2+{dr^2_s\over 1-{2M\over r_s}}+
r^2_s(d\theta^2+sin^2\theta d\phi^2)\end{equation}
Here $r_s$ and $2M$ denote Schwarzschild radial coordinate and mass 
respectively. The junction conditions at the boundary $r=r_b$
are satisfied if $r_s=R_0(T)=R(t,r_b), F(r_b)=2M$.
The tangent vector to the
geodesics in the interior is given by equations (8) and (9)
for radial null geodesics. The tangent vector to radial null
geodesics in the Schwarzschild exterior are
\begin{equation}K^T={k\over 1-{2M_s\over r}},\quad K^r =k\end{equation}
where $k$ is constant and is determined by the continuity of
the tangent vectors to null geodesics in the interior and the exterior
at the boundary
\begin{equation}k={1-{F(r_b)\over R(t,r_b)}\over \sqrt{1+f(r_b)}+
\sqrt{f(r_b)+
{F(r_b)}\over R(t,r_b)}}[k^t]_{r=r_b}\end{equation}
Using equations (36) to (38) we get for the redshift as observed
by the distant observer at $r_s=r_1$ with four velocity 
$U_o^a=(\delta^a_T/ \sqrt{1-{2M\over r_1}})$ in the Schwarzschild exterior
receiving photons which terminate
in the past at the naked central singularity as
\begin{equation}1+z = {(1-{{2M\over r_1}})^{1/2}\over 
1-\sqrt{{2M\over R_o}}}I
\end{equation}
\begin{equation}I=\exp{(\int_{0}^{r_b}Ndr)}=
\exp{(\int_{0}^{R_b}N{dr\over dR}dR)}=
\exp{(\int_{0}^{R_b}{dR\over R-F(r)}q(R,r))}\end{equation}
where we have put $f=0$ and 
\begin{equation}q(R,r)=\sqrt{{\Lambda\over X}}(1+\sqrt{{\Lambda\over X}})
{\Theta -2(1-\Theta r^{\beta})X^{{3\over 2}}\over
\Theta +(1-\Theta r^{\beta})X^{{3\over 2}}}\end{equation}
value of $q$ between the interval $(0,R_c)$ is finite 
and our interest is only in cases when $R_b$ is close to $F(r_b)=2M$
the main contribution to the integral $I$ comes from the factor $R-F$ in the
denominator, and the contribution of this term at the upper limit $R_b$
very close to $F(r_b)$ is most significant. 
We therefore have after careful consideration the redshift behavior
as
\begin{equation}1+z\propto (1-{2M\over R_o})^{n-1}
\propto e^{{T_o(1-n)\over 4m}}\end{equation}
where constant $n$ depends on the $r_b,F(r_b)$ and we have used geodesic
equations$^1$ (37) in Schwarzschild spacetime to express the result in 
terms of the observer time $T_o$. The point to note
is that even the radiations from naked singularity in the late stages of
collapse are redshifted which increase exponentially with a different
characteristic time constant compared with the characteristic time for
radiations coming from the surface of the star given in (1). 
Therefore the radiations from the singularity
would not necessarily disappear simultaneously with the  
radiations from the surface of the star in such cases.

{\bf References}

1. M. A. Podurets, Astr. Zh. 41, 1090 (1964); I. H. Dwivedi and R. Kantowski
Lecture Notes in Physics, Springer-Verlag, 14, 127 (1970)

2. Wagh, B and Lake K.  Phys. Rev D 38 4, 1315(1988);
K. Lake  Phys. Rev. Lett.68, 3129 (1992).

3. R.P.A.C. Newman Class. Quantum Grav. $\underline
{3}$ 527 (1986); ; D. Christodoulou, Commun Math. Phys. 93 , 171 (1984);

4. A.Ori and T. Piran, Phys. Rev. D 42, 1068 (1990)
I H Dwivedi and P S Joshi Commun. Math. Phys. 146 333(1992)

5. I H Dwivedi and P S Joshi Phys. Rev. D 45, 6, 2147(1992);
I H Dwivedi and P S Joshi Class. and Quantum Grav. 9, L69 (1992);

6. I H Dwivedi and P S Joshi Phys. Rev. D 47, 12, 5357(1993)

7. Tolman R. C. Proc. Natl. Acad Sci USA 20 410 (1934)

8. Jhingan S. and Joshi P. S. Annals of the Israel Physical Society {\bf 13},
375(1998)

9. The weak singularity here means in the sense of Tipler, but they are
strong in the sense of Krolak. It should however be noted that the 
singularities satisfying the Krolak condition also may not be extended 
through. Furthermore, a recent study (Deshingkar S., Dwivedi I. H. and
Joshi P. S.; to appear) has shown that even the so called weak singularities 
in dust collapse are strong in the sense of Tipler as well, although they 
have a directional behavior as far as the strength of the singularity is 
concerned. Therefore, the strong sense here should be taken more in terms 
of the growth of density from outer layers towards the center.

\end{document}